\documentclass[twocolumn,aps,showpacs,pre]{revtex4}
\usepackage{graphicx}% Include figure files
\usepackage{verbatim}

\begin{document}
\title{The tensor renormalization group study of the general spin-S Blume-Capel model}
\author{Li-Ping Yang$^{1,2}$}\email{liping2012@cqu.edu.cn}
\author{Zhi-Yuan Xie$^3$}
\affiliation{$^{1}$Department of Physics, Chongqing University, Chongqing 401331, China}
\affiliation{$^{2}$Beijing Computational Science Research Center, Beijing, 100094, China}
\affiliation{$^{3}$Institute of Physics, Chinese Academy of Sciences, P.O. Box 603, Beijing 100190, China}
\begin{abstract}
We focus on the special situation of $D=2J$ of the general spin-S Blume-Capel model on the square lattice.
Under the infinitesimal external magnetic field, the phase transition behaviors due to the thermal fluctuations are discussed by the newly developed tensor renormalization group method.
For the case of the integer spin-S, the system will undergo $S$ first-order phase transitions with the successive symmetry breaking
with the magnetization $M=S,S-1,...0$. For the half-integer spin-S, there are similar $S-1/2$ first order phase transition
with $M=S,S-1,...1/2$ stepwise structure, in addition, there is a continuous phase transition due to the spin-flip $Z_2$ symmetry breaking.
In the low temperature regions, all first-order phase transitions are accompanied by the successive disappearance of the optional spin-component pairs($s,-s$), furthermore, the critical temperature for the nth first-order phase transition is the same, independent of the value of the spin-S. In the absence of the magnetic field, the visualization parameter characterizing the intrinsic degeneracy of the different phases clearly demonstrates the phase transition process.
\end{abstract}
\maketitle
The Blume-Capel model\cite{BC} and the extended Blume-Emergy-Griffiths model\cite{BEG} have attracted the general interests in the several decades.
The Monte-Carlo algorithm\cite{ PhysRevE.73.036702, PhysRevB.57.11575, pre92}, conformal invariance\cite{PhysRevB.57.11575},
 finite-size scaling\cite{PhysRevB.33.1717}
and mean-field approximation\cite{William,Plascak} demonstrate the rich phase diagrams of this model. The richness
 originates, not only from the spin-flip $Z_2$ symmetry breaking, but also from the density fluctuations with
 $S+1$(for integer $S$) or $S+1/2$(for half-integer $S$) optional values\cite{William} for $S_i ^2$. Furthermore,
  It is a characterization for the experimental 3He-4He mixtures\cite{He1,He2} and metamagnets\cite{metamag},
  which inspired the vector version of this model\cite{vector1,vector2}. The introduction of the bond
  randomness\cite{Malakis1,Malakis2} enriches the phase transition discussions about this model.

The Hamiltonian of the Blume-Capel model\cite{BC} with general spin-S is
\begin{equation}
H=-J\sum_{<i,j>}S_iS_j + D\sum_i S_i^2-h\sum_i S_i^z,\label{eq:Hamil}
\end{equation}
where the spin variable $S$ takes $2S+1$ values: $(-S,-S+1,...,S-1,S)$. $J$ is the coupling constant, $D$ is the strength of the single-ion anisotropy and $h$ is the magnetic field. The sum of the first term runs over all the nearest neighbors. Firstly, we consider the case of $h=0$.
When $D=0$, this model is reduced to the classical Ising model. When $D$ goes to the positive infinity, the energy favorable state is the one with the full occupation of the smallest spin component. For the integer spin cases, the component is $S=0$. For the half-integer cases, the component is $S=\pm 1/2$.

If we look at the hamiltonian of any bond linking two sites $i,j$, then we have
\begin{equation}
H_{ij}=-JS_iS_j+D(S_i^2+S_j^2)/q.
\end{equation}
Here, $q$ is the coordination number depending on the lattice structure.
This formula can be rewritten as
\begin{equation}
H_{ij}=J(S_i-S_j)^2/2+(D/q-J/2)(S_i^2+S_j^2),
\end{equation}
from which, it leads to the following conclusions for the ferromagnetic coupling($J>0$): when $D>qJ/2$,
 the configuration of the ground state is $S_i=S_j=0$, $S_i=S_j=1/2(-1/2)$(depending on the spontaneous breaking),
 respectively for the integer and half-odd spin-S cases;  when $D<qJ/2$, the configuration of the ground state is $S_i=S_j=\max(S)$ for any spin-S cases. As a result, we have a special situation, i.e., $D=qJ/2$, where the ground state is $S_i=S_j$ with $(2S+1)$-fold degeneracy.
Once the magnetic field $h$ is turned on as a smallness, the ground state degeneracy is lifted in the case of $D=qJ/2$. The positive smallness $h$ make the system go into the ground state $S_i=S_j=\max(S)$ as the case $D<qJ/2$ with $h=0$.

Without loss of the generality, we focus on the square lattice hereafter. Then, $q=4$ and $D=2J$ is the special parameter. The phase boundaries of the square lattice\cite{Plascak,PhysRevE.73.036702,PhysRevB.33.1717,PhysRevB.57.11575} end at $(T/J, D/J)=(0,2)$, i.e., for $D=2J$, and the critical temperature is $T_c=0$ for the general spin cases when $h=0$.

Motivated by the special parameter point and the thought how the phase boundary approaches the ending point, we calculate the temperature dependence behaviors of this model described by Eq.~(\ref{eq:Hamil}) by the recently developed tensor renormalization group algorithm\cite{zhiyuan}. This work
aims to demonstrate and visualize the phase transition process by the common and special physical quantities calculated by this algorithm.

\begin{figure}[tbp]
\includegraphics[width=0.8\columnwidth]{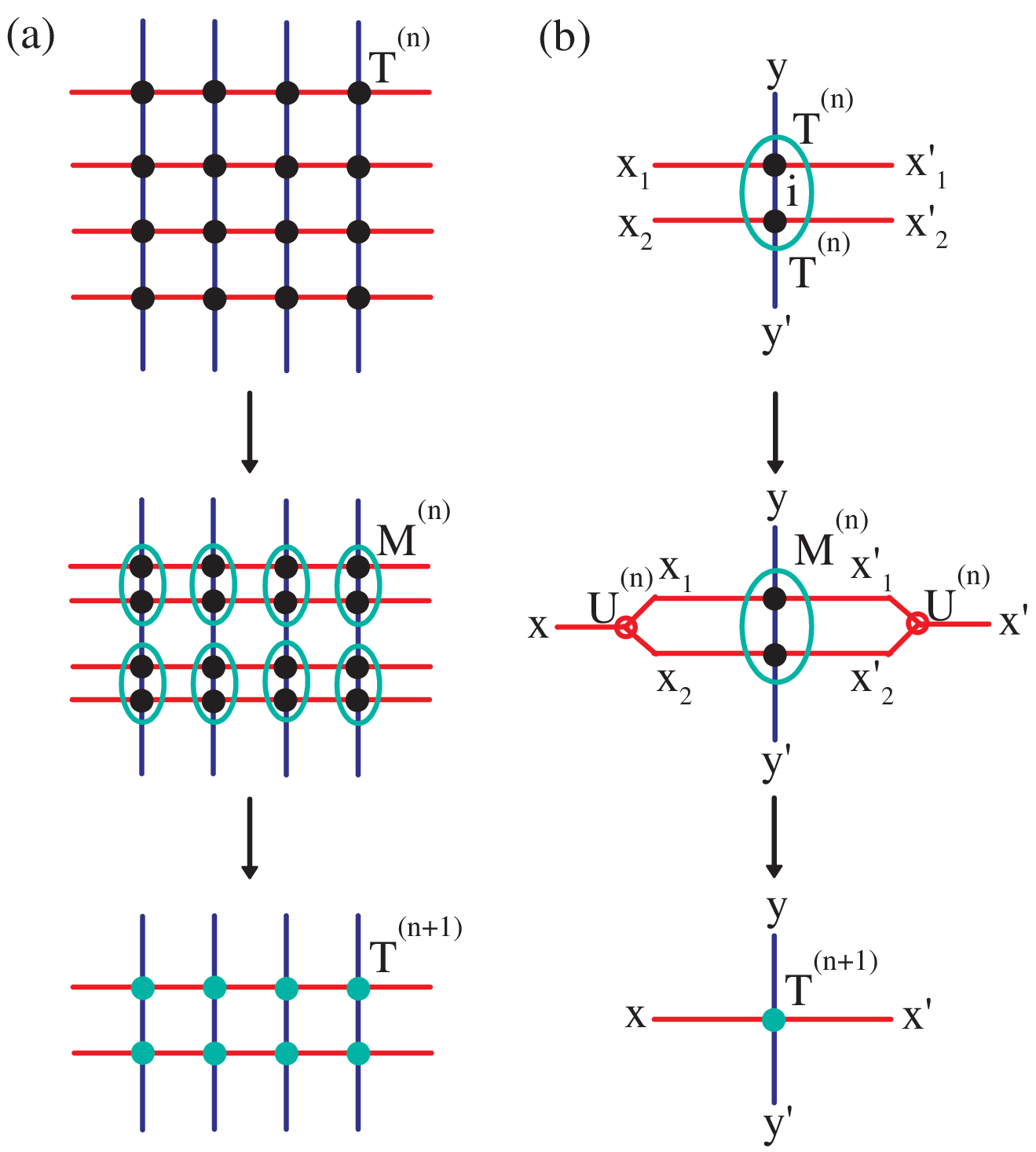}
\caption{
(a) A HOTRG contraction of the tensor network state along the y axis on the square lattice. (b) Steps of contraction and renormalization of two local tensors. The initial tensor $T^{(0)} = T$.
}
\label{fig:SquareRGflow}
\end{figure}

For a classical lattice model with the local interactions, the partition function can be represented as the tensor product\cite{SRG-PRB},
\begin{equation}
Z = \mathrm{Tr}\prod_{i} T_{ x_{i}x'_{i}y_{i}y'_{i} } ,    \label{eq:PartFunc}
\end{equation}
where $i$ runs over all the lattice sites and $\mathrm{Tr}$ is to sum over all bond indices. The local tensor $T$ is defined at each lattice site as shown in Fig.~\ref{fig:SquareRGflow}(a). We have
\begin{equation}
T_{ x_{i}x'_{i}y_{i}y'_{i} }=\sum_{\alpha} W_{\alpha x_{i}} W_{\alpha x'_{i}} W_{\alpha y_{i}} W_{\alpha y'_{i}}.
\end{equation}
$W$ comes from the decomposition of the bond matrix $A_{S_iS_j}=e^{-\beta H_{S_iS_j}}$, i.e., $A=WW^{\dagger}$.
Here,
\begin{equation}
H_{S_iS_j}=-JS_iS_j+D(S_i^2+S_j^2)/q-h(S_i+S_j)/q,
\end{equation}
and the matrix dimension is the spin degree of freedom $2S+1$.

To contract the tensor network, i.e., trace over all sites, we will face the exponential increasing problem of
 the dimension of each order of the tensor. The bond dimensions of the $n$-th and $n-1$-th contraction satisfy $d_n=d_{n-1}^2$. The contraction process will make the dimension inaccessible quickly. In 2007,
 Levin and Nave\cite{Levin} proposed a cutting-off scheme by the singular value decomposition(SVD), keeping an affordable cutting-off dimension $d$ to get a good approximation for the partition function, by which the thermodynamic properties of the system can be obtained consequently.

Recently, we proposed a novel coarse graining tensor network renormalization group(TRG) based on the high order singular value decomposition(HOSVD)\cite{zhiyuan}, abbreviated as HOTRG, which provides an accurate but low computational
cost technique for studying two- or three-dimensional (3D) lattice models. The coarse-graining procedure consists in iteratively replacing
 blocks of size two by a single site using HOSVD along the horizontal (x-axis) and vertical (y-axis) directions alternatively.

 This scheme of coarse graining is simple to be implemented. Fig.~\ref{fig:SquareRGflow}(a), as an example, shows how the contraction along the y-axis is done. At each step, two sites are contracted into a single site in the coarse grained lattice (Fig.~\ref{fig:SquareRGflow}(b)), and the lattice size is reduced by a factor of 2. Therefore, for the system of size $2^{N_s}$, we can obtain the partition function by $N_s$ contraction.

The contracted tensor at each coarse grained lattice site is defined by
\begin{equation}
M^{(n)}_{xx'yy'} = \sum_{i} T^{(n)}_{x_{1}x'_{1}yi} T^{(n)}_{x_{2}x'_{2}iy'} ,  \label{eq:2DFatTensor}
\end{equation}
where $x=x_1\otimes x_2$, $x'=x'_1 \otimes x'_2$, and the superscript $n$ denotes the $n$'th iteration. The bond dimension of $M^{(n)}$ along the x-axis is the square of the corresponding bond dimension of $T^{(n)}$. To truncate $M^{(n)}$ into a lower rank tensor, we first do a HOSVD for this tensor\cite{HOSVD-method}
\begin{equation}
M^{(n)}_{xx'yy'} = \sum_{ijkl} S^{c}_{ijkl}U^L_{xi} U^R_{x'j} U^U_{yk} U^D_{y'l} , \label{eq:2DHOSVD}
\end{equation}
where $U$'s are the unitary matrices. $S^c$ is the core tensor of $M^{(n)}$, which possesses the following properties for any index, say index $j$:
(1) all orthogonality,
\[
\langle S^{c}_{:,j,:,:} \mid S^{c}_{:,j',:,:} \rangle = 0, \qquad \mathrm{if} \,\, j\neq j',
\]
where $\langle S^{c}_{:,j,:,:} \mid S^{c}_{:,j',:,:} \rangle$ is the inner-product of these two sub-tensors. (2) pseudo-diagonal,
\[
|S^{c}_{:,j,:,:}| \geq |S^{c}_{:,j',:,:}|, \qquad \mathrm{if} \,\, j < j',
\]
where $|S^{c}_{:,j,:,:}|$ is the norm of this sub-tensor which is the square root of all elements' square sum. These norms play a similar role as the singular values of a matrix.

In $M^{(n)}$, the two vertical bonds, $y$ and $y'$, do not need to be renormalized. Moreover, the right bond of $M^{(n)}$ is linked directly to the left bond of an identical tensor on the right neighboring site, thus to truncate any one of the horizontal bonds of $M^{(n)}$ will automatically truncate the other horizontal bond. The truncation can be done by comparing the values of $\varepsilon_1 = \sum_{i>d} |S^{c}(i,:,:,:)|^2$ and $\varepsilon_2 = \sum_{j>d} |S^{c}(:,j,:,:)|^2$. If $\varepsilon_1 < \varepsilon_2$, we truncate the first dimension of $S^{c}$ or the second dimension of $U^L$ to $d$. Otherwise, we truncate the second dimension of $S^c$ or the second dimension of $U^R$ to $d$. This provides a nearly optimal approximation to minimize the truncation error. This kind of truncation schemes has in fact been successfully applied to many fields such as data compression, image processing, pattern recognition, and etc \cite{HOSVD-AdvanApp}.

After the truncation, we can update the local tensor using the following formula
\begin{eqnarray}
T^{(n+1)}_{xx'yy'} = \sum_{ij} U^{(n)}_{ix} M^{(n)}_{ijyy'} U^{(n)}_{jx'} ,                                   \label{eq:2DTruncation}
\end{eqnarray}
where $U^{(n)} = U^L$ (or $U^R$) if $\varepsilon_1$ is smaller (or larger) than $\varepsilon_2$.
The above HOTRG calculation can be repeated iteratively until the free energy and other physical quantities calculated are converged. The cost of the calculation scales as $d^7$ in the computer time and $d^4$ in the memory space. This is comparable with the cost of TRG \cite{Levin,SRG-PRB,zhiyuanprl}.

The following results are all from the newly developed HOTRG scheme. Hereafter, the coupling constant $J$ is used as the energy unit, and $k_B$ is set as $1$. The calculation about the magnetization and occupation number, $h$ is taken as $10^{-10}$ in Eq.~(\ref{eq:Hamil}) for
the preferential symmetry breaking of the spin pairs $(s,-s)$ and the computational stability, which makes the effective
single-ion anisotropy parameter $D$ slightly smaller than the
special value $2J$. So the discussion about the magnetization and occupation number corresponds to the left vicinity of the special parameter $D(=2J)$ in the case of $h=0$.

The system size
is fixed at $2^{40}$ for the different spin value cases as shown below. However, $h$ is fixed at $0$ for the visualization parameter\cite{PhysRevB.80.155131} referring to
the state degeneracy, which will be addressed in Sec.~\ref{sec:X1}. The periodic boundary condition is adopted in all the numerical calculation.

As follows, we will discuss the results in the typical integer cases of $S=1,2$ in Section.~\ref{Sec:integer} and half-odd cases of $S=3/2,5/2$ in Sec.~\ref{Sec:half} respectively. Then we analyze
the visualization parameter of the phase transition in Sec.~\ref{sec:X1}. Finally, we go to the conclusion.

\section{the integer cases: $S=1,2$}
\label{Sec:integer}
In the case of $S=1$, there are three optional values: $\pm 1, 0$ for each spin variable.
There exists the tricritical point enriching the phase diagram.

\begin{figure}
\includegraphics[width=1.0\columnwidth]
{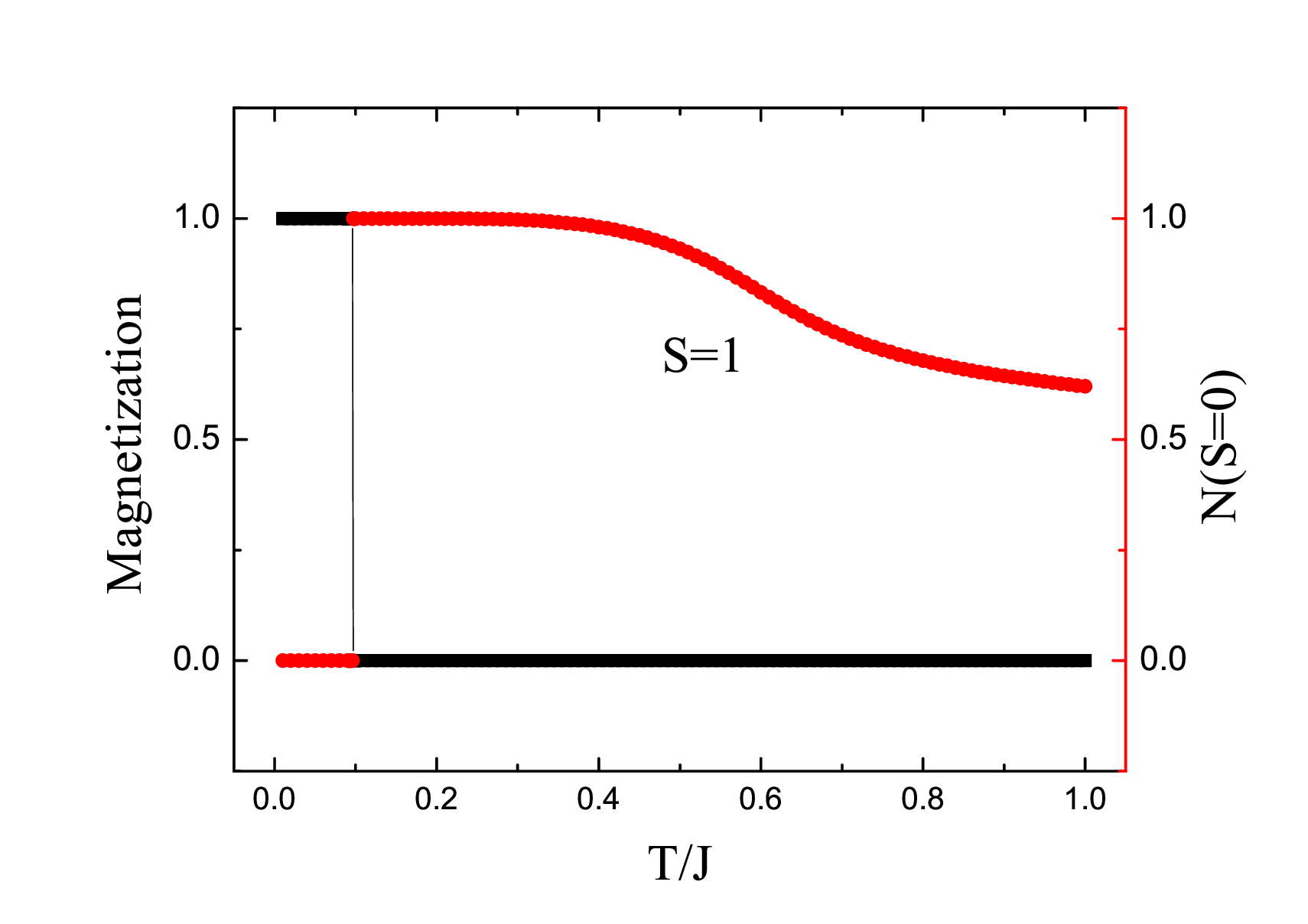} \caption{(Color Online)The temperature dependence of the magnetization(black square) and occupation number of
$N(S=0)$(red circle, light grey), here, $S=1, D=2J, N_s=40, d=30, h=10^{-10}$.} \label{fig:MNS1}
\end{figure}

  In the limit of infinite $D$, the energy favorable state is full occupation of $S=0$, denoted as $N(S=0)=1$ per site.
    We can name $S=0$ as the vacancy or hole, the state full of holes is just the hole condensed phase,
    whose quantum correspondence is the one-dimensional $S=1$ quantum model with single-ion anisotropy
    of our previous papers\cite{PhysRevLett.100.067203,PhysRevB.79.214427}. The hole condensed states also
    emerge in the other general integer spin-S cases.

  As is shown in Fig.~\ref{fig:MNS1}, the system undergoes the first-order phase transitions
  accompanying the emergence of the hole condense phase as the temperature varies.
  The magnetization jumps from $1$ to $0$, whose location is consistent with the kink of
  the occupation number of the pairs $N(S=1,-1)$. The critical point reads $T_c=0.096$. For the terrace of the magnetization $M=0$,
   the starting point is the condensation of the holes with the disappearance of the pair $S=1,-1$. With the temperature further increasing,
     the components $N(S=1,-1)$ come back to the system again with the equal-weight occupation, and the system goes into
     the paramagnetic state with the magnetization $M=0$. In the limit of the high temperature, the occupation numbers of the three different components
     $S=1,0,-1$ all go to $1/3$.

For the case of $S=2$, there are five optional values: $0,\pm 1,\pm 2$ for the spin variable. Compared to the case of $S=1$, there is one more optional pair $S=\pm2$, which renders one more jump of the magnetization from $M=2$ to $M=1$. The system undergoes the first order phase transitions twice with the temperature increasing.

 The first platform $M=2$ corresponds to the full occupation of $N(S=2)=1$ from the breaking of the pair $S=\pm 2$ due to the tiny magnetic field $h$.
The second platform $M=1$ emerges with the disappearance of $N(S=2,-2)$ and full occupation of $N(S=1)=1$ from the breaking of the pair $S=\pm 1$. The second jump of the magnetization comes with the hole condensed phase $N(S=0)=0$. With the temperature further increasing, the occupation of the hole drops down continuously. The degree of the freedom $S=\pm 1$ and $S=\pm 2$ come back to the system gradually. $N(S=0)+N(S=1,-1)\simeq 1$ keeps until the considerable temperature. Accompanying the successive disappearance of $S=\pm2$ and $S=\pm1$, the two critical points are located at $T_{c_1}=0.096$ and $T_{c_2}=0.305$ respectively.

\begin{figure}
\includegraphics[width=1.0\columnwidth]
{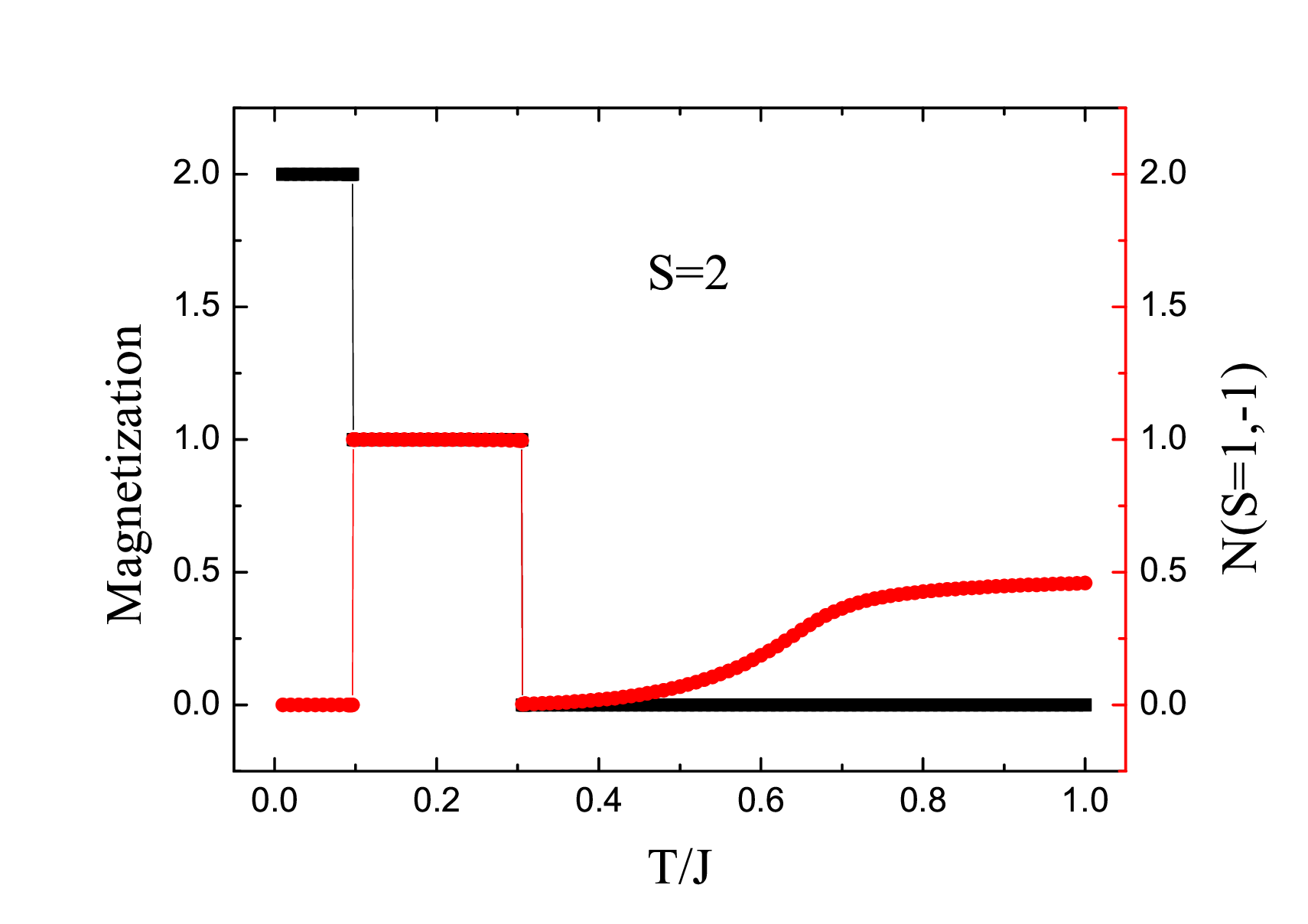} \caption{(Color Online)The magnetization(black square) and occupation number of $N(S=1,-1)$(red circle, light grey) as a function of the reduced temperature $T/J$, here, $S=2, D=2J, N_s=40, d=30, h=10^{-10}$.} \label{fig:MNS2}
\end{figure}
  \section{The half-odd cases: $S=3/2,5/2$}
  \label{Sec:half}
For $S=3/2$, there are four optional values :
$\pm 3/2, \pm 1/2$ for each spin variable.

Similar to the case of $S=1$, the system breaks into the state with $M=3/2$ due to the tiny $h$. The first jump from $M=3/2$ to $M=1/2$ exhibits the competition of the occupation of two pairs between $S=\pm 3/2$ and $S=\pm 1/2$. The fisrt-order phase transition brings the full occupation change from $N(S=3/2)$ to $N(S=1/2)$.

When the temperature increases further, the Ising-like continuous phase transition shows up due to spin-flip $Z_2$ symmetry breaking.
By the behaviors of the magnetization, the two critical points are located at $T_{c1}=0.096$ and $T_{c2}=0.649$ respectively.
 The location of $T_{c2}$ is close to the data shown in Fig.~2 in Ref.\onlinecite{PhysRevB.57.11575}.
\begin{figure}
\includegraphics[width=1.0\columnwidth]
{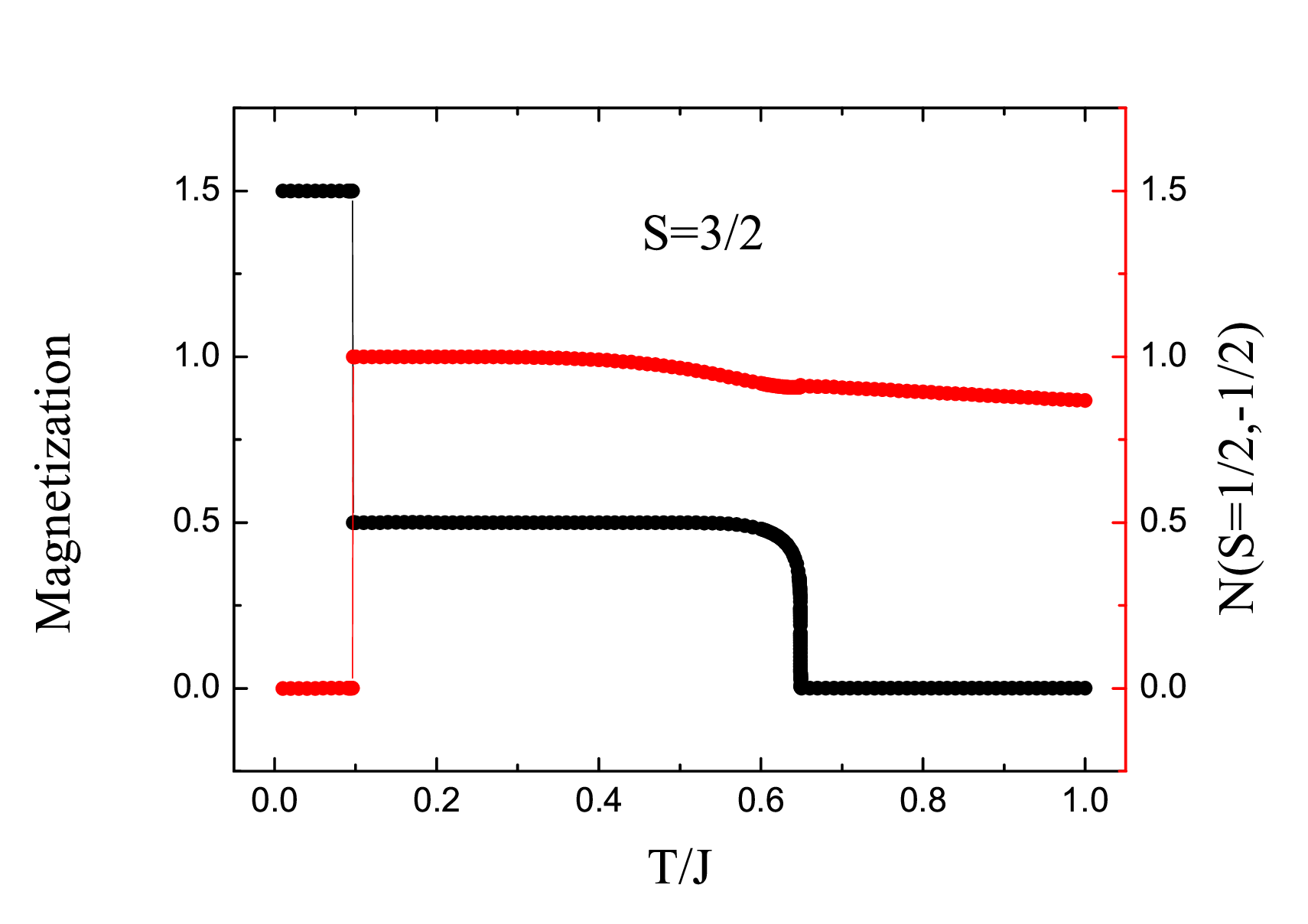} \caption{(Color Online)The temperature dependence of the magnetization(black square) and occupation number
$N(S=1/2,-1/2)$(red circle, light grey), here, $S=3/2, D=2J, N_s=40, d=40, h=10^{-10}$.} \label{fig:MNS1d5}
\end{figure}

\begin{figure}
\includegraphics[width=1.0\columnwidth]
{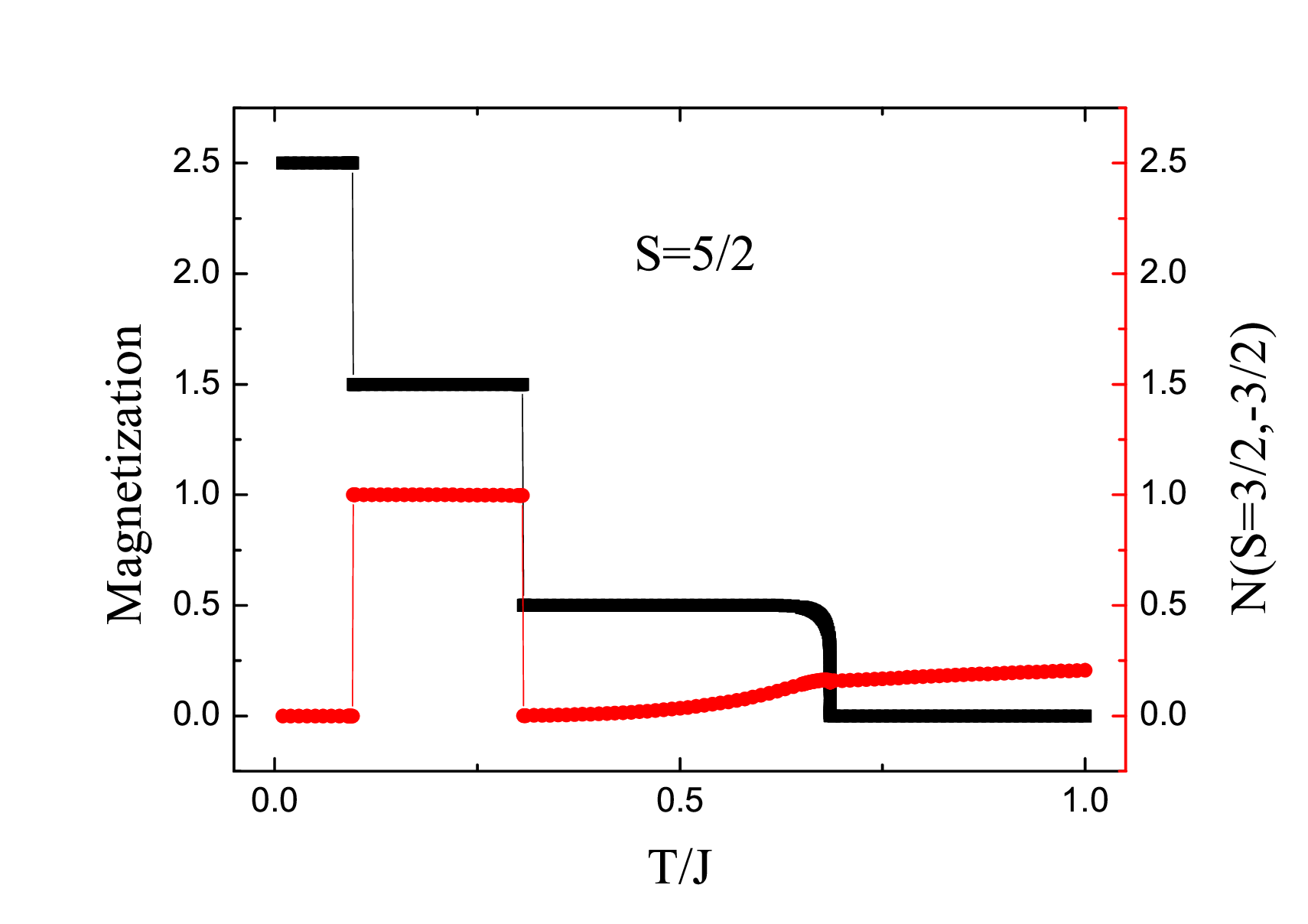} \caption{(Color Online)The temperature dependence of the magnetization(black square) and occupation number $N(S=3/2,-3/2)$(red circle, light grey), here, $S=5/2, D=2J, N_s=40, d=40, h=10^{-10}$.} \label{fig:NhS25}
\end{figure}

The similar discussion is applied to the case of $S=5/2$, where there are six optional values: $\pm 1/2, \pm 3/2, \pm 5/2$ for each spin variable.

The first two jumps in the magnetization are the phase transitions of the first-order. The successive symmetry breaking of the pairs $(s,-s)$ is clearly shown in the
profile of the occupation number $N(S=3/2,-3/2)$. At the first terrace $M=5/2$,
the degrees of freedom $S=(3/2,-3/2),(1/2,-1/2)$ are gone. On the way to increase the temperature,
the first jump of the magnetization $M$ and the occupation number $N$ indicates the disappearance
 of $N(S=5/2,-5/2)$ and full occupation of $N(S=3/2,-3/2)$. Then the second jump of $M$ and $N$ represents the ensuing
   replacement of $N=(S=3/2,-3/2)$ by $N=(S=1/2,-1/2)$. The symmetry breaking of the
    spin pair $S=(1/2,-1/2)$ due to $h$ brings $N(S=1/2)=1$.

  However, with the temperature further increasing, $M$ drops to $0$ continuously, accompanying
   the continuous phase transition, and the system enters into the paramagnetic phase. Here,
   the occupation number $N(S=3/2,-3/2)$ increases continuously from $0$. A further calculation
    shows that $N(S=1/2)=N(S=-1/2), N(S=3/2)=N(S=-3/2)$ in the paramagnetic phase,
     where $N(S=1/2,-1/2)+N(S=3/2,-3/2)\simeq 1$ keeps up to the considerable temperature
     until the equal-weight distribution of all the degree of freedom in the high temperature limit.
     The three critical points read $T_{c_1}=0.096, T_{c_2}=0.306, T_{c_3}=0.685$ successively.

Reviewing the data about the critical points, we find that the critical temperature for the 1st
 first-order phase transition is fixed at $T_{c_1}=0.096$ for the above four cases $S=1,3/2,2,5/2$.
 And the 2nd fist-order phase transition is located at $T_{c_2}=0.305(0.306)$ for $S=2(5/2)$.

Let us turn to the principle about the minimal free energy, $F=U-TS$. The phase transition lies in the competition between
the internal energy and the entropy. Assuming that there are two clusters filled with single spin component $s, s+1$ separately, the existence of the interface connecting the two clusters will raise the internal energy.
Simultaneously the entropy is increased due to more possible configurations. The 1st first-order phase transition occurs with the global spin-component replacement of $S$ by $S-1$ in each site, and the transition point is irrelevant of the value of spin-S. The similar discussion applies to nth first-order phase transition in the low temperature situation. The same interval of spin components brings the same internal
   energy difference. For the low temperature regime, the relative Boltzmann weight $e^{-\beta\Delta E}$ due to the variation of the spin components is negligible. As a consequence, we can not see the appreciable thermal fluctuations, then the magnetization plateau emerges.

 The additional check shows that position of the 1st first-order phase transition moves closer to zero
 temperature with smaller $h$. However, the second-order phase transition points are insensitive with smallness $h$. In the relatively high temperature, the effect from the tiny $h$ is negligible. Our results about the continuous phase transition points $T_c=0.649,0.685$ for $S=3/2,5/2$ respectively will be the reference data. All 1st first-order phase transitions will occur at zero temperature for the exact $D=2J, h=0$ situation , which is a spontaneous symmetry breaking picture in the thermodynamic limit and exactly demonstrated in Ref.~\onlinecite{Plascak}.

     Nevertheless,
      the strength of the smallness of $h$ doesn't change the fact that
      the corresponding location of the nth first-order
      phase transition keep the same, independent of the spin values in the low temperature regime.
        Although the tiny magnetic field $h$ shifts the critical point along the temperature axis.
       The perspective from the
       visualization parameter\cite{PhysRevB.80.155131} also provides the qualitative consistence.

\section{visualization parameter}
\label{sec:X1}
As was pointed out in Ref.~\onlinecite{PhysRevB.80.155131}, a symmetry breaking phase with $n$ degenerate states is represented by fix-point
 tensors, which is a direct sum of $n$ dimension-one trivial tensor.  $X_{1}$ is introduced to visualize the structure of fixed-point tensors
  with the
 following definition,
\begin{equation}
X_{1}=\frac{(\sum_{ru}T_{ruru})^2}{\sum_{ruld}T_{rulu}T_{ldrd}}.
\end{equation}
It is independent of the scale of the tensor, and the graphical demonstration is referred to Fig.~13 in Ref.~\onlinecite{PhysRevB.80.155131}.
$X_{1}$ directly
represents the information of the degeneracy associated with the symmetry underlying the Hamiltonian.
\begin{figure}
\includegraphics[width=1.0\columnwidth]
{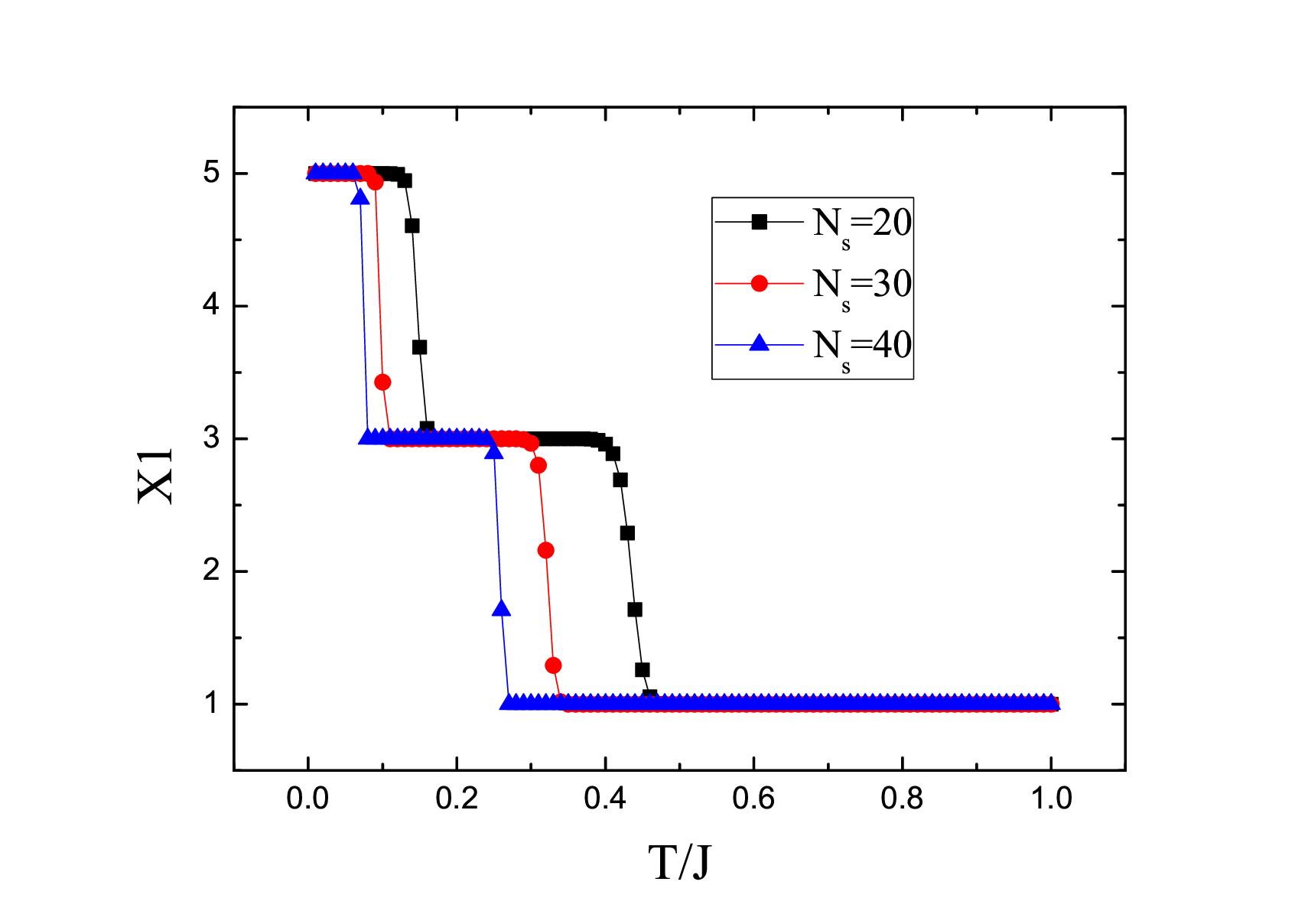} \caption{(Color Online)The step-wise structure of the visualization $X_{1}$ with three platforms $(5,3,1)$ as a function of the reduced temperature $T/J$. Three different lattice sizes $N_s=20,30,40$ are compared. Here, $S=2, D=2J, d=40, h=0$.} \label{fig:X1S2}
\end{figure}

For the integer case of $S=2$, the visualization $X_{1}$ bears the step-wise structure: $5,3,1$. The $5$-fold degeneracy associated
to $T^{Z_2}\bigoplus T^{Z_2}\bigoplus T^{TRI}$ is represented by the first plateau valued $5$.
 $T^{Z_2}$ corresponds to the spin flip of the pair $(s,-s)$ and $T^{TRI}$ comes from the degree of freedom $S=0$.
 With the temperature increasing, the system undergoes two phase transitions. The first jump of $X_{1}$ from $5$ to $3$ and $T^{Z_2}$
 originating from $S=\pm 2$ is gone, where $N(S=1)$ exhibits the step change from $0$ to $1$. The plateau of $X_{1}=3$
 corresponds to $T^{Z_2}\bigoplus T^{TRI}$ referring to $S=0,1,-1$. The plateau of $X_{1}=1$ is the
 consequent vanishing of $T_{Z_2}$ about another pair $S=\pm 1$, where $N(S=0)=1$.

\begin{figure}
\includegraphics[width=1.0\columnwidth]
{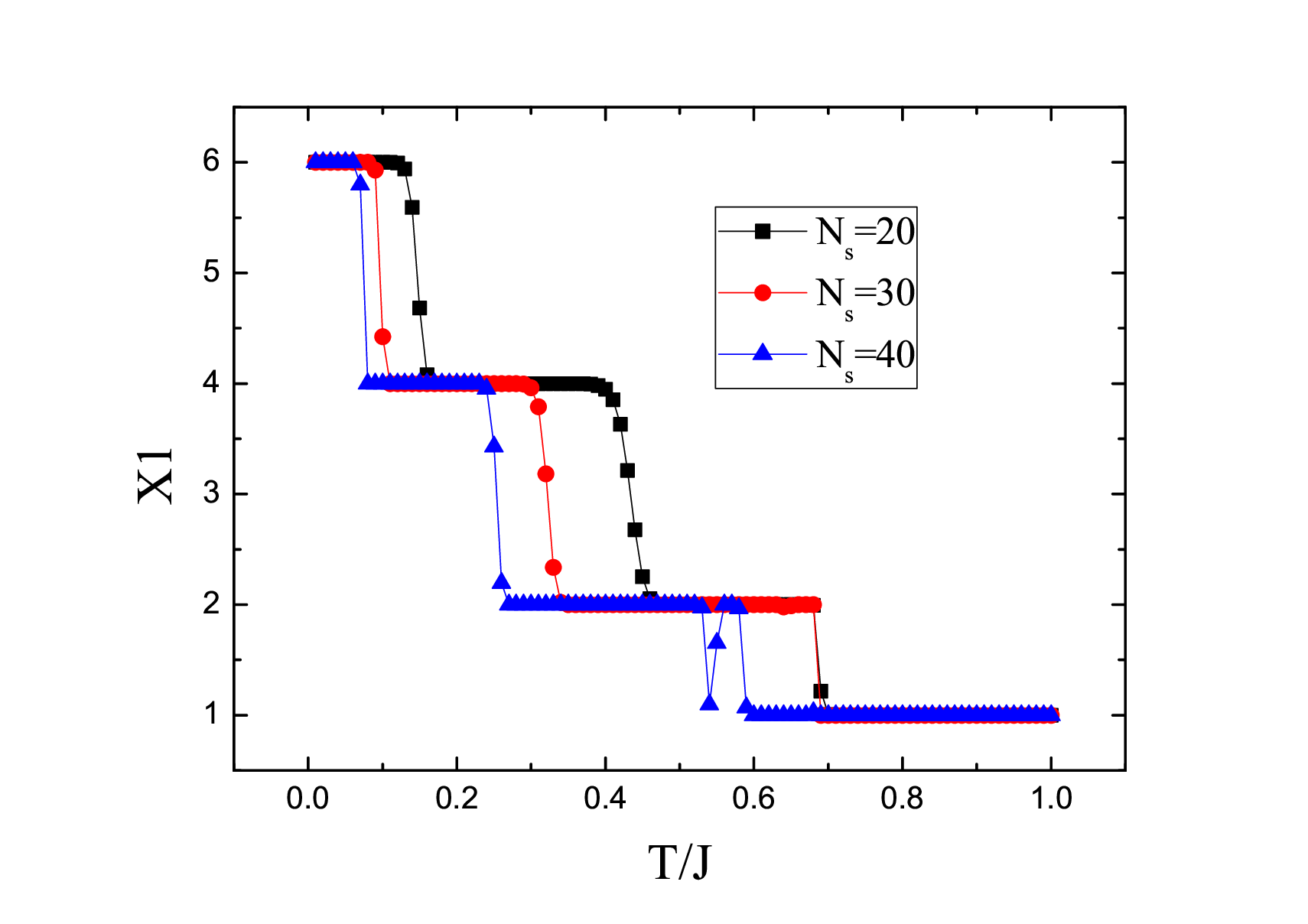} \caption{(Color Online)The step-wise structure of the visualization $X_{1}$ with four platforms $(6,4,2,1)$ as a function of the reduced temperature $T/J$. Three different lattice sizes $N_s=20,30,40$ are compared. Here, $S=5/2, D=2J, d=40, h=0$.} \label{fig:X1S25}
\end{figure}

 $X_{1}$ is used to characterize the degeneracy, which is the reason why $h$ is fixed at $0$ here. The introduction of $h$ will lift the degeneracy.
 And the tiny numerical calculation error will affect the numerical results about the intrinsic degeneracy, we choose the different system sizes $2^{N_s}, N_s=20,30,40$ for comparison. With $N_s$ increasing, the transition points move to the left, corresponding to the lower transition temperature points. The first transition point should be $T_c=0$ when $D=2J,h=0$.
 On the other side, the big system size will bring numerical instability for $X1$ with more coarse-graining steps, as shown in Fig.~(\ref{fig:X1S25}).
 When the symmetry describing the state degeneracy are not implemented in the construction of the original tensor, we will face the problem. The research about the fix-point tensor is still on the way.

For the half-odd case of $S=5/2$, the visualization $X_{1}$ bears the step-wise structure: $6,4,2,1$. The $6$-fold degeneracy comes from
 the three pairs $\pm 5/2, \pm 3/2, \pm 1/2$, corresponding to $T^{Z_2}\bigoplus T^{Z_2}\bigoplus T^{Z_2}$. The first two jumps of $X_{1}$
 represent the two first order phase transitions, which are similar to the case of $S=2$ accompanying the successive vanishing of
  the occupation pairs. The difference lies in the last jump of $X_{1}$ from $2$ to $1$, which is the continuous phase transition with $Z_2$
   spin-flip symmetry breaking. The gap of $X_{1}$ is $2$ for all the first order phase transitions and $1$ for the last continuous
   Ising-like phase transition respectively.

It leads to the general conclusion that the visualization parameter $X_{1}$ bears the step-wise structure: $2S+1,2S-1,...,3,1$ for the
 integer spin $S$, $2S+1,2S-1,...,2,1$ for the half-odd spin $S$. It is tricky to locate the position of the critical point by the jump
  of $X_{1}$ because of the subtlety of $X_{1}$ due to the cutting-off in the coarse-graining process, however, the integer plateaus of $X_{1}$ associated to the degeneracy of the system is intrinsic, which comes from the deep insight that the fix-point of the tensor representation corresponds to the fix point of RG flow. It provides an reference to observe the phase transition. By the implement of the symmetry in the initial tensor construction, the phase transition point of
    ferromagnetic potts models on the simple cubic lattice identified by $X_{1}$ can beat the most recent Monte Carlo
result\cite{wangshun}.

In summary, we discuss the phase transitions of the Blume-capel model on the square lattice using the recently developed HOTRG.
$h=0$, $D=2J$ is a special situation with the high energy degeneracy, where the bond Hamiltonian is the form of the square sum. When $h$ is turned on as a smallness with the same magnitude,
  We find that the location of the nth first-order phase transition
   is the same independent of the spin-S in the low temperature regions, because the difference of the adjacent spin component is always $1$.
   The position of the successive phase transitions can be identified more exactly by increasing the cutting-off dimension $d$. Through the magnetization and the occupation number of different optional spin values, the first-order phase transition behaviors
    are clearly demonstrated. The visualization parameter $X_{1}$ associated to the
    degeneracy illustrates the successive phase transitions. Compared to the integer spin-S cases, there is one
    more Ising-like continuous phase transition with spin-flip $Z_2$ symmetry breaking for half-odd spin-S cases.

We would like to thank Xiao-Gang Wen, Tao Xiang, Bruce Normand, You-Jin Deng, Xiao-Yong Feng, Ming-Pu Qin and Jing Chen for the stimulating
discussions. The work was supported by Natural Science Foundation of China for the Youth (Grants No.11304404) and partly supported by MOST (Grant No. 2011CB922204).
%\bibliography{references}

\begin{thebibliography}{99}
\bibitem{BC} M. Blume, Phys. Rev. \textbf{141}, 517 (1966); H. W. Capel, Physica \textbf{32}, 966 (1966).
\bibitem{BEG} M. Blume, V. J. Emergy and R. B. Griffiths, Phys. Rev. A \textbf{4}, 1071 (1971).
\bibitem{PhysRevE.73.036702} C. J. Silva, A. A. Caparica, and J. A. Plascak, Phys. Rev. E \textbf{73}, 036702 (2006).
\bibitem{PhysRevB.57.11575} J. C. Xavier, F. C. Alcaraz, D. Pen\~a Lara, and  J. A. Plascak, Phys. Rev. B \textbf{57}, 11575 (1998).
\bibitem{pre92} W. Kwak, J. Jeong, J. Lee, D.-H. Kim, Phys. Rev. E \textbf{92}, 022134 (2015)
\bibitem{PhysRevB.33.1717} Paul D. Beale, Phys. Rev. B \textbf{33}, 1717 (1986).
\bibitem{William} W. Hoston and A. N. Berker, Phys. Rev. Lett. \textbf{67}, 1027 (1991).
\bibitem{Plascak} J. A. Plascak, J. G. Moreira, and F. C. S\'a Barreto, Phys. Lett. A \textbf{173}, 360 (1993).
\bibitem{He1} E. H. Graf, D. M. Lee, and J. D. Reppy, Phys. Rev. Lett. \textbf{19}, 417 (1967).
\bibitem{He2} G. Goellner and H. Meyer, Phys. Rev. Lett. \textbf{26}, 1534 (1971).
\bibitem{metamag} V. A. Schmidt and S. A. Friedberg, Phys. Rev. B \textbf{1}, 2250 (1970).
\bibitem{vector1} J. L. Cardy and D. J. Scalapino, Phys. Rev. B \textbf{19}, 1428 (1979).
\bibitem{vector2} A. N. Berker and D. R. Nelson, Phys. Rev. B \textbf{19}, 2488 (1979).
\bibitem{Malakis1} A. Malakis, A. Nihat Berker, I. A. Hadjiagapiou and N. G. Fytas, Phys. Rev. E \textbf{79}, 011125 (2009).
\bibitem{Malakis2} A. Malakis, A. Nihat Berker, I. A. Hadjiagapiou and N. G. Fytas and T. Papakonstantinou, Phys. Rev. E \textbf{81}, 041113 (2010).
\bibitem{zhiyuan} Z. Y. Xie, J. Chen, M. P. Qin, J. W. Zhu, L. P. Yang, and T. Xiang,  Phys. Rev. B \textbf{86}, 045139 (2012).
\bibitem{Levin} M. Levin and C. P. Nave, Phys. Rev. Lett. \textbf{99}, 120601 (2007).
\bibitem{HOSVD-method} L. de Latheauwer, B. de Moor, and J. Vandewalle, SIAM J. Matrix Anal. Appl, \textbf{21}, 1253 (2000).
\bibitem{HOSVD-AdvanApp} D. J. Luo, C. Ding, and H. Huang, arXiv:0902.4521; G. Bergqvist, E. G. Larsson, IEEE Signal Proc. Mag. \textbf{27}, 151 (2010).
\bibitem{SRG-PRB} H. H. Zhao, Z. Y. Xie, Q. N. Chen, Z. C. Wei, J. W. Cai, and T. Xiang, Phys. Rev. B \textbf{81}, 174411 (2010).
\bibitem{zhiyuanprl} Z. Y. Xie, H. C. Jiang, Q. N. Chen, Z. Y. Weng, and T. Xiang, Phys. Rev. Lett. \textbf{103}, 160601 (2009).
\bibitem{PhysRevB.80.155131} Z.-C. Gu,  and X. -G. Wen, Phys. Rev. B \textbf{80}, 155131 (2009).
\bibitem{PhysRevLett.100.067203} Z. Yang, L. Yang, J. Dai, and T. Xiang, Phys. Rev. Lett. \textbf{100}, 067203 (2008).
\bibitem{PhysRevB.79.214427} Z. H. Yang, L. P. Yang, H. N. Wu, J. Dai, and T. Xiang, Phys. Rev. B \textbf{79}, 214427 (2009).
\bibitem{wangshun}S. Wang, Z. Y. Xie, J. Chen, N. Bruce, T. Xiang, Chin. Phys. Lett \textbf{31}, 070503 (2014)




\end{thebibliography}

\end{document}